\documentclass[fleqn,twoside]{article}
\usepackage{espcrc2}

\newcommand{\be}{\begin{equation}}
\newcommand{\ee}{\end{equation}}

\def\bea{\begin{eqnarray}}
\def\eea{\end{eqnarray}}

\def\jrn#1#2#3#4{{#1} {\bf #2} (#4) #3}
\def\PRL{\it Phys. Rev. Lett.}
\def\PLB{\it Phys. Lett. B}

\def\NPB{\it Nucl. Phys. B}

\newcommand{\AmS}{{\protect\the\textfont2
  A\kern-.1667em\lower.5ex\hbox{M}\kern-.125emS}}

\title{Neutrinos: Key to New Physics}

\author{P. Ramond\address{Institute for Fundamental Theory, Physics Department,   
        University of Florida, Gainesville, FL 32611, United States}%
        \thanks{This Research is supported in part by 
the US Department of Energy under grant DE-FG02-97ER41029.}
             }
       
\begin{document}

\begin{abstract}
The Seesaw mechanism  predicted tiny neutrino masses by postulating a new large scale in particle physics, using new theoretical ideas prompted by the Standard Model. It adds credence to a theoretical vista that is a quarter century old, and fits with the most endearing speculations of ultimate unification. By relating the measurement of static neutrino properties  to  near-Planck physics, it may even prove key to solving the riddles of flavor. 
\vspace{1pc}
\end{abstract}

\maketitle

\section{INTRODUCTION}
Twenty five years ago,  the Seesaw Mechanism\cite{SEESAW} was born in the afterglow of the intellectual turmoil generated by the Standard Model. The renormalizability of massive Yang-Mills theories\cite{HOOFT}, the emergence  of a common description of Weak and Electromagnetic Interations\cite{STANDARD}, and the realization that the Strong Interactions  weaken at shorter distances\cite{AF} established the Standard Model as the paradigm for all Fundamental Interactions except Gravity. Like all models of Nature, no matter how successfull, it is incomplete,  has generated new puzzles, and elicited many questions. None has been more dominating than Pati and Salam's\cite{PS} proposal that quarks and leptons are equal partners of one mathematical structure at very short distances, the so-called  Grand-Unification. 

No celebration of the Seesaw mechanism would be complete without couching it in the context of a bread-brush presentation of the great theoretical speculations of its time. Below, we begin the  weaving of such a  tapestry.

\section{STANDARD MODEL LEGACIES}
The Standard Model describes  fundamental interactions down to millifermis. Rather than describing it in detail we merely note some of its startling features:

\noindent -- The Standard Model has emerged practically unscathed from almost four decades of experiments. Measurements of the $Z$-boson width  puts a limit of three active neutrinos, a decade of precision measurements have vindicated its radiative structure, and all of its quarks and leptons have been discovered. One  parameter remains to be fixed by experiment, the mass of  the elusive Higgs particle.  One prediction of the Standard Model has been put to rest by experiments:  neutrinos have masses.  

\noindent -- Interactions stem from  three {\it weakly coupled } Yang-Mills theories based on $SU(3)$, $SU(2)$ and $U(1)$. 

\noindent --  Both quarks and leptons are needed for quantum consistency:  gauge anomalies cancel between quarks and leptons.

\noindent --  There are three chiral families of quarks and leptons, each with a massless neutrino.

 \noindent -- The gauge symmetries are spontaneously broken: the  shorter the distance, the {\it more} the symmetry.

\noindent --  It predicts a fundamental scalar particle, the Higgs boson.

\section{ NEW \& OLD PUZZLES }
Although the Standard Model has been successfull beyond  expectations, there is a dark side to it, with one new puzzle,   some unsavory features, and some wrong predictions:

\noindent --  It predicts CP-violation in the Strong interaction, albeit with unknown strength.

\noindent --  It requires  Yukawa interactions but does not offer any organizing principle.

 \noindent -- It fails to {\it explain} the values of the  masses and mixing patterns of  quarks and charged leptons. 

\noindent --  It  contains too many parameters.

\noindent --  It  is not a complete description of Nature since it does not address gravitational interactions. It only describes only the matter side of Einstein's equation, {\em sans}  cosmological constant.
 
\noindent -- It fails to account for neutrino masses.

For these and other reasons, it is obvious that the Standard Model is an unfinished picture that needs to be put in a more general framework where  these shortcomings are addressed. It appears to be  like the shards of a once beautifull pottery, shattered in the course of cosmological evolution.

\section{ GRAND UNIFICATION}
The quantum numbers within each of the three families of quarks and leptons strongly suggest a more unified picture. It is remarkable that Pati and Salam's original idea is realized by  unifying  the three gauge groups of the Standard Model into one simple gauge group. In the simplest\cite{GG}, $SU(5)$, each family appears in two representations. In $SO(10)$\cite{FM}, they are grouped  in the fundamental spinor representation, by adding a right-handed neutrino for each family. At the next level of complication, we find $E_6$\cite{ESIX} where each family contains several right-handed neutrinos as well as vector-like matter. Organizing the elementary particles into these beautiful structures    

 \noindent -- Unifies  three gauge groups  into one.

\noindent -- Relates quarks and Leptons.

\noindent -- Explains anomaly cancellations.

There are indications that this idea ``wants to work". When last seen, the three coupling constants of the Standard Model are perturbative. Using the renormalization group equations to continue them  deep into the ultraviolet, they  get closer to one another, but fail to meet at one scale. Originally, the Weinberg angle was not measured with sufficient accuracy, and it was thought that they met at one point, suggesting a new scale of physics, close to the Planck mass: the quantum number patterns did not quite match the dynamical information. This near unification introduced Planck scale physics into the realm of particle physics.

A by-product of this Grand Unification is the  violation of baryon number. Hitherto unobserved, proton decay remains one of the most important consequences 
from these ideas. In a serendipitous twist, proton decay detectors now  serve as the telescopes of neutrino astronomy! Other global symmetries also bite the dust: the relative lepton numbers are violated in $SU(5)$ and $SO(10)$ violated the total lepton number as well, and the extraordinary limits on these processes are  consistent with  the grand-unified scale.

\section{GRAND-UNIFIED LEGACIES}
Although Grand Unification by itself does not yet have any direct experimental vindication, it has proven to be an incubator of ideas that, to this day, drive  speculations on the Physics of extra-short distances. 

\noindent --  It showed how the large grand-unified scale could be used to generate tiny neutrino masses\cite{SEESAW}.

\noindent -- It suggested some links between quark and charged lepton masses, with some success for the two heaviest families. Still, the flavor riddles of the Standard Model remain largely unexplained by Grand Unification.

\noindent -- It  created the ``gauge hierarchy" problem,  why quantum corrections leave the ratio of the  Higgs mass  to the Grand-Unified  scale unscathed. 

Moreover,  two of its predictions have linked particle physics to pre-Nucleosynthesis Cosmology:

\noindent -- It predicted the existence of monopoles in our universe.  This led to the idea of Inflationary Cosmology\cite{GUTH},  which  solves many long standing puzzles and whose prediction of a flat universe has been recently verified experimentally. 

\noindent -- It predicted proton decay, and offered a framework for understand the baryon asymmetry\cite{YOSHIMURA} of the Universe.   

Today, only one of these predictions, tiny neutrino masses, has been borne out by experiment. Like the Standard Model, it clearly is not a complete theory of Nature, since it does not address Gravity (space-time is either flat or a fixed background ), nor the origin of the three chiral families and the associated flavor puzzles.

\section{ SUPERSTRINGS}
 At the 1973 London conference, David Olive  in his rapporteur talk, declared  Superstring Theories to be ``Theories of Everything". As he stated, 
Superstring theories  reproduce Einstein's gravity at large distances with no ultraviolet divergences, and can also contain (some) gauge theories. This view has since gained much credence and notoriety. The  matter content has gotten much closer to reality\cite{HET}, although this unification of the gravitational and gauge forces takes place in a somewhat unsettling background:

\noindent -- Fermions and Bosons are related by a new type of symmetry: supersymmetry!      

\noindent -- Ultimate Unification takes place in more than three space dimensions!

Nature at the millifermis displays neither supersymmetry nor extra space dimensions. Yet,  the lesson of the Standard Model of more symmetries at shorter distances provide an argument for these to be fabrics of the Ultimate Theory; they are shattered by cosmological evolution.  To compare the highly symmetric superstring theories to Nature, a dynamical understanding of the breakdown of these symmetries is required, an understanding that  still eludes us. 

To make contact with experiments, it is necessary to know the energy at which these symmetries manifest themsemselves. Both types could be  just around the energy corner, but  I would like to argue that   circumstantial evidence lends more credence to low- energy supersymmetry than to low-energy extra dimensions. For some reason,  the collapse of the extra space dimensions occurs first, while Supersymmetry hangs on to later times (lower energies). It is a challenge to theory to find a dynamical reason which triggers the breakdown of higher-dimensional space (perhaps through  brane formation),  while leave supersymmetry nearly intact.


\section{ SUPERSYMMETRY}
 Supersymmetry is clearly an attractive theoretical concept; it  is required by the unification of gravity and gauge interactions, and links fermions and bosons. Also, the mass of the spinless superpartner of a Weyl fermion, inherits quantum-naturality \cite{GILDENER} through  the chiral symmetry of its partner.  
 
Morever, when applied to the Standard Model, it yields quantitative predictions that fit remarkably well with Gauge Unification. 

\noindent --   The Gauge hierarchy problem is managed: the mass of the Higgs is stabilized even in the presence of a large (grand-unification) scale

\noindent --  The three gauge couplings of the Standard Model run to a single value in the deep ultraviolet with the addition of superpartners in the TeV range. Thus naturally emerges a new scale  using the renormalization group, a scale that matches the quantum number patterns of the elementary particles. 

\noindent -- With supersymmetry the renormalization group displays an infrared fixed point that predicts\cite{PENDLETON} a heavy the top quark, in agrrement with experiment.

\noindent -- Under a  large class of ultraviolet initial conditions, the same renormalization group shows that the breaking of supersymmetry  triggers electroweak breaking\cite{EWBREAK}. 

Supersymmetry at low energy is the leading theory for physics beyond the Standard Model, although  many puzzles remain unanswered and  new ones are created as well. 

Firstly, there are almost as many theories of  supersymmetry breaking as there are theorists, and  none, theories and theorists alike,  are convincing. It is an experimental question.

Secondly it adds little to the flavor problem; rather it makes it worse by  predicting  new scalar particles which generically produce flavor-changing neutral processes. Even if the breaking mechanism is flavor-blind (tasteless), non-trivial effects are expected: supersymmetry-breaking is already highly constrained by the existing data set.

All will be forgotten when superpartners are discovered  at the LHC.  May the supersymmetry- breaking mechanism parameters prove to be  bizarre enough to allow intellectually-challenged theorists  to infer its origin from the LHC data alone!

\section{ TINY NEUTRINO MASSES}
 The only solid experimental evidence to date for physics beyond the Standard Model is the observation of oscillation among neutrino species.  Experiments on solar neutrinos
\cite{Homestake,SKsol,SNO} yield 

 $$\Delta m^2_\odot~=~\vert~m^2_{\nu_1}-m^2_{\nu_2}~\vert~\sim~7.\times 10^{-5}_{}~{\rm eV^2}\ ,$$
with corroborating evidence on antineutrinos\cite{Kamland}. Neutrinos born in Cosmic ray collisions\cite{SKatm}, and on earth\cite{K2K} give
     
$$\Delta m^2_\oplus~=~\vert~m^2_{\nu_2}-m^2_{\nu_3}~\vert~\sim~3.\times 10^{-3}_{}~{\rm eV^2}\ .$$
The best bound to their absolute value of the masses comes from WMAP\cite{WMAP}

$$\sum_i~m^{}_{\nu_i}~<~.71~{\rm eV}\ .$$
These experimental findings are not sufficient to determine fully the mass patterns. One oscillates between  three patterns, {\it hierarchy}, 

$$|m_{\nu _1}| < |m_{\nu _2}| \ll |m_{\nu _3}|\ ,$$
{\it  inverse hierarchy}

$$|m_{\nu _1}| 
\simeq |m_{\nu _2}| \gg |m_{\nu _3}|\ ,$$
or even {\it hyperfine}

$$|m_{\nu _1}|  \simeq |m_{\nu _2}| \simeq |m_{\nu _3}|\ .$$
The mixing patterns provide some surprises, since it contains one small angle and two large angles. In terms of the MNS mixing matrix,

$$\pmatrix{\cos\theta^{}_\odot&\sin\theta^{}_\odot&\epsilon\cr
-\cos\theta^{}_\oplus~\sin\theta^{}_\odot&\cos\theta^{}_\oplus~\cos\theta^{}_\odot&\sin\theta^{}_\oplus\cr
\sin\theta^{}_\oplus~\sin\theta^{}_\odot&-\sin\theta^{}_\oplus~\cos\theta^{}_\odot&\cos\theta^{}_\oplus}\ ,$$
the various experiments yield
     
$$ \sin^2 2\theta^{}_\oplus~>~0.85\ ,\qquad  0.30~<~ \tan^2\theta^{}_\odot~< ~0.65 \ ,$$
while there is a only a limit\cite{CHOOZ} on the third angle 

$$ \vert~\epsilon~\vert^2_{}~<~0.05\ .$$
Spectacular as they are, these results generate new questions for experimenters:

\noindent -- Are the neutrino masses Majorana-like (i.e. lepton number violating)?

\noindent -- What is their absolute values? Can one measure the sign of $\Delta m^2$?

\noindent -- Is  CP-violation in the lepton sector observable?

\noindent They also generate new theoretical questions

\noindent -- Are there right-handed neutrinos? 

\noindent -- If so, how many, how heavy, with what hierarchy? 

\noindent --  Where do they live? Brane or bulk? 

\noindent -- Do they cause leptogenesis?

\section{ Standard Model Analysis}
Masses and mixings of the quarks are determined from the diagonalization of Yukawa matrices generated by the  $\Delta I_{\rm W}=\frac{1}{2}$ breaking 
of electroweak symmetry, for charge $2/3$ 

$$
{\cal U}^{}_{2/3}\,
\pmatrix{m^{}_u&0&0\cr 0&m^{}_c&0\cr 0&0&m_t^{}}
\,{\cal V}^{\dagger}_{2/3}\ ,$$
and charge $-1/3$

$$
{\cal U}^{}_{-1/3}\,
\pmatrix{m^{}_d&0&0\cr 0&m^{}_s&0\cr 0&0&m_b^{}}
\,{\cal V}^{\dagger}_{-1/3}\ ,
$$
resulting in the observable CKM matrix

$$
{\cal U}^{}_{CKM}~\equiv~{\cal U}^{\dagger}_{2/3}\,{\cal U}^{}_{-1/3}\ .
$$
Up to Cabibbo-size effects, it is equal to the unit matrix, implying that mixing is similar for up-like and down-like quarks. Their masses are of course highly hierarchical.

The charged lepton Yukawa matrix  

$${\cal U}^{}_{-1}\,
\pmatrix{m^{}_e&0&0\cr 0&m^{}_\mu&0\cr 0&0&m_\tau^{}}
\,{\cal V}^{\dagger}_{-1}$$
also stems from $\Delta I_{\rm W}=\frac{1}{2}$ electroweak breaking. To generate neutrino masses,  add  one right-handed neutrino for each  family, producing its own Yukawa matrix

$${\cal U}^{}_{0}\,\pmatrix{m^{}_1&0&0\cr 0&m^{}_2&0\cr 0&0&m_3^{}}\,{\cal V}^{\dagger}_{0}\ .$$
In order to proceed,  the nature of the right-handed neutrino's masses needs to be specified.  They are  of the Majorana type. Since the right-handed neutrinos have no gauge quantum numbers, their masses necessarily violate total  lepton number.

In the spirit of effective field theories, one therefore expects their masses to be of the order of lepton number breaking. Total lepton number-violating processes have never been seen resulting in a bound from neutrinoless double $\beta$ decay experiments. So either they are very large or zero. If they are zero, the analysis is like that in the quark sector, and the observable MNS lepton mixing matrix is just

$${\cal U}^{}_{MNS}~\equiv~ {\cal U}^{\dagger}_{-1}\,{\cal U}^{}_{0}\ .$$
It would be generated solely from the isospinor breaking of electroweak symmetry, just like the quarks', even though the mixing patterns are so different. 

In the belief that global symmetries are an endangered species (black holes eat them up),  we expect their masses to set the scale of the Standard model's cut-off, since they are unprotected by gauge symmetries. This yields the Seesaw where large right-handed neutrino masses engender tiny neutrino masses, the latter being suppresses over that of the charged particles by the ratio of the two scales
     
$$\frac{\Delta I_{\rm W}=\frac{1}{2}}{\Delta I_{\rm W}=0}\ .$$
This introduces a large electroweak-singlet scale   in the Standard Model.  The neutrino mass matrix is then  
 
$${\cal M}^{(0)}_{   {Seesaw}}~=~{\cal M}^{(0)}_{  {Dirac}}\,
\frac{1}{{\cal M}^{(0)}_{   {Majorana}}}\,{\cal M}^{(0)\,T}_{  {Dirac}}\ ,$$
which we can rewrite as

$${\cal M}^{(0)}_{   {Seesaw}}~=~{\cal U}^{}_{0}\,\,
  {\bf{\cal C}}\,\,{\cal U}^{T}_{0}\ ,$$
in terms of the central matrix\cite{DLR}

$$ {\cal C}~=~{\cal D}_0^{}\,{\cal V}^{\dagger}_{0}\,\frac{1}{{\cal M}^{(0)}_{   {Majorana}}}\,
{\cal V}^{*}_{0}\,{\cal D}_0^{}\ .$$
It is diagonalized by the unitary matrix ${\cal F}$
 
$$~~~  {\cal C}~=~   {\cal F}\,{\cal D}^{}_\nu\,   {\cal F^{\,T}_{}}\ ,$$
where the mass eigenstates produced in $\beta$-decay are (unimaginatively labelled  as ``1", ``2", ``3")      

$$
{\cal D}_\nu^{}~=~\pmatrix{m^{}_{\nu_1}&0&0\cr 0&m^{}_{\nu_2}&0\cr 0&0&m_{\nu_3}^{}}\ .$$
The effect of the seesaw  is to add the unitary ${\cal F}$ matrix to the MNS lepton matrix 

$${\cal U}^{}_{MNS}~=~ {\cal U}^{\dagger}_{-1}\,{\cal U}^{}_{0}\,\,   {\cal F}\ .$$
This framework enables us to recast our theoretical questions in terms of $\cal F$. So we can ask where the large angles reside: it is convenient to catalog the models in terms of the number of large angles contained in  $\cal F$, none, one or two? 

\section{A Modicum of  Grand Unification}  
To relate the CKM and MNS matrices and the quark and lepton masses, the natural framework is of course grand unification. There,  the  $\Delta I^{}_{\rm W}=\frac{1}{2}$  quark and lepton Yukawa matrices  are related, using the simplest Higgs contents. 

At the level of $SU(5)$, the charge $-1/3$ and charge $-1$ Yukawa matrices are family-transposes of one another. 

$$ {\cal M}^{(-1/3)}_{}~\sim~{\cal M}^{(-1)\,T}_{}\ .$$
In $SO(10)$, it is the charge $2/3$ Yukawa matrix that is related to the Dirac charge $0$ matrix

 $$ {\cal M}^{(2/3)}_{}~\sim~{\cal M}^{(0)}_{  {Dirac}}\ .$$
These result in naive expectations for the unitary matrices that yield observable mixings

$${\cal U}^{}_{-1/3}~\sim~{\cal V}^{*}_{-1}\ ;\qquad {\cal U}^{}_{2/3}~\sim~{\cal U}^{}_{0}\ .$$
Assuming these, we can relate the CKM and MNS matrices

\bea\nonumber{\cal U}^{}_{MNS}&=& {\cal U}^{\dagger}_{-1}\,{\cal U}^{}_{0}\,   {\cal F}\cr 
 \nonumber&\sim&{\cal U}^{\dagger}_{-1}\,{\cal U}^{}_{-1/3}\,{\cal U}^{\dagger}_{CKM}\,   {\cal F} \cr
&\sim&     \Big({{\cal V}^T_{-1/3}\,{\cal U}^{}_{-1/3}}\,\Big)\,{\cal U}^{\dagger}_{CKM}\,\,   {\cal F}\eea
  Hence two wide classes of models:

\noindent I-) Models with Family-Symmetric ${\cal M}^{}_{-1/3}$ Yukawa matrices. 
In these we have  

$$    {{\cal U}^{}_{-1/3}}~=~    {{\cal V}^*_{-1/3}}\ ,$$
so that      

$$
{  {{\cal U}^{}_{MNS}~=~ {\cal U}^{\dagger}_{CKM}\,\,   {\cal F}
}}\ .$$
In these models, $   {\cal F}$    necessarily contains two large angles. In the absence of any symmetry acting on $\cal F$, these models appear to be of a type I call non-generic. In particular they could require a non-Abelian structure for $\cal F$.
\vskip .3cm

\noindent II-) Models with Family-Asymmetric ${\cal M}^{}_{-1/3}$ Yukawa matrices.  If we extend the Wolfenstein\cite{WOLF} expansion of the CKM matrix in powers of the Cabibbo angle $\lambda$ to include quark mass ratios

$$ \frac{m^{}_s}{m^{}_b}~\sim~   {\lambda^2_{}}\qquad \frac{m^{}_d}{m^{}_b}~\sim~   {\lambda^4_{}}\ ,$$
we find the charge $-1/3$ Yukawa matrix
 
$$    {{\cal M}^{(-1/3)}_{}}~=~  {\pmatrix{   {\lambda^4_{}}&   {\lambda^3_{}}&   {\lambda^3_{}}\cr
    {\lambda^?_{}}&   {\lambda^2_{}}&   {\lambda^2_{}}\cr
    {\lambda^?_{}}&    {\lambda^{?}_{}}&{1}}}\ .$$
If the exponents are related to charges, as in the Froggatt-Nielsen\cite{FN} schemes, then the lower diagonal exponents are known, and we get the orders of magnitude
 
$$    {{\cal M}^{(-1/3)}_{}}~=~  {
\pmatrix{   {\lambda^4_{}}&   {\lambda^3_{}}&   {\lambda^3_{}}\cr
    {\lambda^3_{}}&   {\lambda^2_{}}&   {\lambda^2_{}}\cr
    {\lambda^1_{}}&    1&1}}\ ,$$
which is hardly family symmetric. In the limit of no Cabibbo mixing, 

$$    {{\cal M}^{(-1/3)}_{}}~\approx~  {\pmatrix{0&0&0\cr 0&0&0\cr 0&  {a}&  {b}}}
+{\cal O}({   \lambda})\ ,$$
 and 

$${\cal U}^{}_{MNS}~=~
\pmatrix{1&0&0\cr 0&\cos\theta^{}_\oplus&\sin\theta^{}_\oplus\cr 0&-\sin\theta^{}_\oplus&\cos\theta^{}_\oplus}\,   {\cal F}\ ,$$
where 
     
$$\tan\theta^{}_\oplus~=~\frac{a}{b}\ , $$
is of order one\cite{ILR}. In these models, $ {\cal F}$ need  contain only one large angle, which is easily accomodated, a more generic alternative. 

Models  type I provide a hint as to the size of the CHOOZ angle. With a symmetric charge $-1/3$ matrix, the MNS matrix reads

\bea
& &{\cal U}^{}_{MNS}~=~{\cal U}^{\dagger}_{CKM}\,\times \cr & &\cr
& &\pmatrix{\cos\theta^{}_\odot&\sin\theta^{}_\odot&   {\lambda^\gamma}\cr
-\cos\theta^{}_\oplus~\sin\theta^{}_\odot&\cos\theta^{}_\oplus~\cos\theta^{}_\odot&\sin\theta^{}_\oplus\cr
\sin\theta^{}_\oplus~\sin\theta^{}_\odot&-\sin\theta^{}_\oplus~\cos\theta^{}_\odot&\cos\theta^{}_\oplus}
\ ,\nonumber\eea
where we have chosen to fill the zero in the $\cal F$ matrix by a Cabibbo efect of unknown order. 
Then it is easy to see that 

$$\theta^{}_{13}~\sim~\cases
{~~   {\lambda^\gamma}~~\cr ~~   {\lambda}~  {\sin\theta_\oplus}~\sim~\frac{1}{\sqrt 2}\,\lambda\ .
}$$
 It will be interesting to see if this precise prediction of type I models is borne out by experiments. 

Models where the charge $-1/3$ Yukawa matrix is not symmetric, no such precise prediction is possible. If we set 

\bea
&&{\cal U}^{}_{MNS}~=~\pmatrix{1&   {\lambda^\alpha}&   {\lambda^\beta}\cr
   {\lambda^\alpha}&
\cos\theta^{}_\oplus&\sin\theta^{}_\oplus\cr    {\lambda^\beta}& -\sin\theta^{}_\oplus&\cos\theta^{}_\oplus}\,\times \cr
&&\cr
&&~~~~~~~~~~~~~~~~~~
~   {\pmatrix{  {\cos\theta^{}_\odot}&  {\sin\theta^{}_\odot}&   {\lambda^\gamma}\cr   {-\sin\theta^{}_\odot}&  {\cos\theta^{}_\odot}&   {\lambda^\delta}\cr    {\lambda^\gamma}&   {\lambda^\delta}&  {1}}}\nonumber\ ,\eea
we see that the CHOOZ angle can take on any number of values $\theta^{}_{13}~\sim~{\lambda^\gamma}\ , ~ {\lambda^{\alpha+\delta}}$, or 
$ {\lambda^\beta}\ $,  depending on the relative values of the exponents. 

Models of either  type suggest a Wolfenstein expansion for the MNS matrix, but the problem is the starting point. Perhaps something like 

$${\cal U}^{}_{MNS}~\sim~ \pmatrix{\cos\alpha &\sin\alpha &0\cr -\frac{\sin\alpha}{\sqrt{2}}&\frac{\cos\alpha}{\sqrt{2}}&\frac{1}{\sqrt{2}}\cr
\frac{\sin\alpha}{\sqrt{2}}&-\frac{\cos\alpha}{\sqrt{2}}&\frac{1}{\sqrt{2}}}~+~{\cal O}(   {\lambda})\ ,$$
with $\alpha=\frac{\pi}{4}$ or $ \frac{\pi}{6}$? Finally we mention that in the quark sector, Cabibbo mixing is strongest between the first and second families. If this effect permeates the leptons, then some {\it Cabibbo flop} might be expected, and one should not put too much value as to their precise values, especially for $1-2$ mixing, so the flop in $\theta_\odot$ could be as much as $\lambda/\sqrt{2}$. These issues and CP-violation will be discussed elsewhere\cite{DER}.

\section{Correlations}
In most models,  $\cal F$ must contain at least one large angle to accomodate the data. This presents a small puzzle since $\cal F$ diagonalizes a matrix which contains the neutral Dirac Yukawa matrix which is presumably hierarchical, coming from the isospinor electroweak breaking. 
This suggests special restrictions put upon the Majorana mass matrix of the right-handed neutrinos. We can see this by looking at a $2\times 2$ two-families case\cite{DLR}.  Let us write

 $$
{\cal D}^{}_0~=~m\pmatrix{a\,   {\lambda^\beta_{}}&0\cr 0&1}\ , $$
and define $M^{}_1\ , M^{}_2$ to be the eigenvalues of the right-handed neutrino's Majorana mass matrix. This matrix can be diagonalized by a large mixing angle in one of two cases:

\noindent -- Its matrix elements have similar orders of magnitude $   {{\cal C}_{11}} ~\sim~    {{\cal C}_{22}} ~\sim~    {{\cal C}_{12}} $, in which case we find that 

$$
\frac{M_1}{M_2}~\sim~   {\lambda^{2\beta}_{}}\ ,$$
suggesting a doubly {\it correlated hierarchy} betwen the $\Delta I_{\rm W}=0$ and $\Delta I_{\rm W}=\frac{1}{2}$ Sectors. This agrees well with grand-unified models such as $SO(10)$ and $E_6$, where each  right-handed neutrinos is part of a family.

\noindent --A large mixing angle can occur if the diagonal elements are much smaller than the diagonal ones, that is 
${\cal C}_{11}\, , \,{\cal C}_{22} ~\ll~ {\cal C}_{12}$. Then we find 

$$\frac{\lambda^\alpha~m^2}{\sqrt{-M_1 M_2}}\,
\pmatrix{0&a\cr a&0}\ .$$
Hence maximal mixing may infer that some of the right-handed neutrinos are Dirac partners of one another, leading to conservation in this matrix of some relative lepton number. 
 
\section{Conclusions}
There is still much to be learned from leptons. With the Seesaw Mechanism, neutrino data can provide a glimpse of physics that can never be reached by accelerators. This  new era of the physics centers around  right-handed neutrinos. With no electroweak quantum numbers, they may hold the key to the flavor puzzles. The second large neutrino mixing angle suggests that hierarchy is independent of electroweak breaking, ans occurs at grand-unified scales. I conclude this talk by noting that Fukugita and Yanagida's wonderful idea of leptogenesis\cite{LEPTOGENESIS} from these neutrinos is much more credible with such a doubled correlated hierarchy. 

 I would like to express my thanks to Professors K. Nakamura,  Y. Totsuka and T. Yanagida for inviting me to  such an intellectually stimulating workshop, and also for  the wonderful hospitality they showed me. I am also grateful to the Fujihara Foundation, Professor Nishijima and the Japan Academy.

\end{document}